\documentstyle[preprint,eqsecnum,aps]{revtex}
\begin{document}
\draft
\title{QUANTIZATION VIA CLASSICAL ORBITS }
\author{De-Hone Lin\thanks{%
e-mail: d793314@phys.nthu.edu.tw}}
\address{department of Physics, National Tsing Hua University \\
Hsinchu 30043, Taiwan}
\date{\today}
\maketitle
\begin{abstract}
A systematic method for calculating higher-order corrections of the
relativistic semiclassical fixed-energy amplitude is given. The central
scheme in computing corrections of all orders is related to a time ordering
operation of an operator involving the Van Vleck determinant. This study
provides us a new viewpoint for quantization.
\end{abstract}
\newpage \tolerance=10000
\section{Correction of All Order for the Semiclassical Relativistic
Fixed-energy Amplitude}
It was pointed out by Van Vleck \cite{0} that the semiclassical
approximation of the propagator in quantum mechanics can be expressed via
the superposition of terms involving the classical action in the exponent,
and allowing for more than one possible classical paths between two
specified points in a given time interval $(t_{b}-t_{a})$: 
\begin{equation}
K_{{\rm sc}}({\bf {x}}_{b},{\bf {x}}_{a};t_{b}-t_{a})=\sum_{{\rm %
classical\;orbit}}\left( \frac{M}{2\pi \hbar i}\right) ^{D/2}\sqrt{\det
\left( -\frac{\partial ^{2}R}{\partial {\bf {x}}_{b}\partial {\bf {x}}_{a}}%
\right) }\exp \left\{ \frac{i}{\hbar }R({\bf {x}}_{b},{\bf {x}}%
_{a};t_{b}-t_{a})\right\} ,  \label{a1}
\end{equation}
where $R$ is the Hamilton principal function, and ${\bf {x}}_{b},{\bf {x}}%
_{a}$ is the terminal points of the orbits. This resulting formula are often
analytically quite complicated, but they have the great merit of describing
almost all the physics. \ Especially, they are often astonishingly accurate;
this is important, because it is precisely in the semiclassical limit that
many of the standard calculational methods of wave mechanics converge very
slowly \cite{0.5}.

In this letter, we would like to present the semiclassical approximation of
the relativistic fixed-energy amplitude (Green's function). The corrections
of all orders is given by a ``time'' like ordering operator. This study
provides us a new viewpoint for quantization.

The starting point is Kleinert's path integral representation of the
fixed-energy amplitude of a relativistic particle in external static
electromagnetic fields \cite{1,2,2.5} 
\begin{equation}
G({\bf x}_{b},{\bf x}_{a};E)=\frac{i\hbar }{2Mc}\int_{0}^{\infty }ds\int 
{\cal D}\rho (\lambda )\Phi \left[ \rho (\lambda )\right] \int {\cal D}%
^{3}x(\lambda )\exp \left\{ -A_{E}/\hbar \right\} \rho (0)  \label{a2}
\end{equation}
with the action 
\begin{equation}
A_{E}=\int_{\lambda _{a}}^{\lambda _{b}}d\lambda \left[ \frac{M}{2\rho
\left( \lambda \right) }{\bf \dot{x}}^{2}\left( \lambda \right) -i(e/c){\bf %
A(x)\cdot \dot{x}(}\lambda {\bf )}-\rho (\lambda )\frac{\left( E-V({\bf x}%
)\right) ^{2}}{2Mc^{2}}+\rho \left( \lambda \right) \frac{Mc^{2}}{2}\right] ,
\label{a3}
\end{equation}
where $s$ in Eq. (\ref{a2}) is defined as 
\begin{equation}
s_{b}-s_{a}=\int_{\lambda _{a}}^{\lambda _{b}}d\lambda \rho (\lambda ),
\label{a31}
\end{equation}
in which $\rho (\lambda )$ is an arbitrary dimensionless fluctuating scale
variable, $\rho (0)$ is the terminal point of the function $\rho (\lambda )$%
, and $\Phi \lbrack \rho (\lambda )]$ is some convenient gauge-fixing
functional [3-6]. The only condition on $\Phi \lbrack \rho (\lambda )]$ is
that 
\begin{equation}
\int {\cal D}\rho (\lambda )\Phi \left[ \rho (\lambda )\right] =1.
\label{a4}
\end{equation}
$\hbar /Mc$ is the well-known Compton wave length of a particle of mass $M$, 
${\bf A(x)}$ is the vector potential, $V({\bf x})$ is the scalar potential, $%
E$ is the system energy, and ${\bf {x}}$ is the spatial part of the ($3+1$)
vector $\vec{x}=({\bf {x}},\tau )$.

It is without lost the generally that the functional $\Phi \left[ \rho
(\lambda )\right] $ is taken \ as the $\delta $-functional $\delta \left[
\rho -1\right] $ to fixed the value of $\rho (\lambda )$ to unity \cite
{1,2,2.5,lin}. Then the lowest order's approximation of the fixed-energy
amplitude for a relativistic system is given by \cite{1,2,2.7} 
\[
G_{{\rm sc}}({\bf {x}}_{b},{\bf {x}}_{a};E)=\frac{\hbar }{2Mc}\frac{1}{(2\pi
\hbar i)^{D/2}}\int_{s_{a}}^{\infty }ds_{b}e^{\left[ \frac{i}{\hbar }{%
\varepsilon }\left( s_{b}-s_{a}\right) \right] }K_{{\rm sc}}({\bf {x}}%
_{b},s_{b}\mid {\bf {x}}_{a},s_{a})
\]
\begin{equation}
=\frac{\hbar }{2Mc}\frac{1}{(2\pi \hbar i)^{D/2}}\int_{s_{a}}^{\infty
}ds_{b}e^{\left[ \frac{i}{\hbar }{\varepsilon }\left( s_{b}-s_{a}\right) %
\right] }D_{E}^{1/2}({\bf {x}}_{b},s_{b}\mid {\bf {x}}_{a},s_{a})e^{\left[ 
\frac{i}{\hbar }{\cal A}_{E}({\bf {x}}_{b},s_{b}\mid {\bf {x}}_{a},s_{a})%
\right] }  \label{sc}
\end{equation}
with the pseudoaction 
\begin{equation}
{\cal A}_{E}=\int_{s_{a}}^{s_{b}}ds\left[ \frac{M}{2}{\bf \dot{x}}^{2}\left(
s\right) +{\frac{e}{c}{\bf A(x,}}s{{\bf )\cdot \dot{x}}(}s)+\frac{1}{2Mc^{2}}%
\left( V^{2}({\bf x},s)-2EV({\bf x,}s)\right) \right] ,
\end{equation}
where $D_{E}$ is the second derivative with respect to ${\cal A}_{E}$ and is
given by 
\begin{equation}
\left| \det \left[ -\partial _{x_{b}^{j}}\partial _{x_{a}^{i}}{\cal A}_{E}(%
{\bf {x}}_{b},s_{b}\mid {\bf {x}}_{a},s_{a})\right] \right| ,
\end{equation}
and the pseudoenergy ${\varepsilon }$ is defined as $%
(E^{2}-M^{2}c^{4})/2Mc^{2}.$ We have assumed that ${\bf A(x,}s{\bf )}$ and $%
V({\bf x},s)$ are functions of coordinate ${\bf x}$ and the timelike
parameter $s$. This will be useful for reducing to non-relativistic
semi-classical approximation.

To get the higher order corrections, we make a reasonable conjecture \cite
{mr} 
\[
G_{{\rm all}}({\bf {x}}_{b},{\bf {x}}_{a};E)=\frac{\hbar }{2Mc}\frac{1}{%
(2\pi \hbar i)^{D/2}}\int_{s_{a}}^{\infty }ds_{b}e^{\left[ \frac{i}{\hbar }{%
\varepsilon }\left( s_{b}-s_{a}\right) \right] }K_{{\rm all}}({\bf {x}}%
_{b},s_{b}\mid {\bf {x}}_{a},s_{a}) 
\]
\[
=\frac{\hbar }{2Mc}\frac{1}{(2\pi \hbar i)^{D/2}}\int_{s_{a}}^{\infty
}ds_{b}e^{\left[ \frac{i}{\hbar }{\varepsilon }\left( s_{b}-s_{a}\right) %
\right] } 
\]
\begin{equation}
\times \left\{ D_{E}^{1/2}({\bf {x}}_{b},s_{b}\mid {\bf {x}}_{a},s_{a})e^{%
\left[ \frac{i}{\hbar }{\cal A}_{E}({\bf {x}}_{b},s_{b}\mid {\bf {x}}%
_{a},s_{a})\right] }\sum_{k=0}^{\infty }\hbar ^{k}g^{(k)}({\bf {x}}%
_{b},s_{b}\mid {\bf {x}}_{a},s_{a})\right\} .  \label{2}
\end{equation}
The subscript ``{\it all''}{\rm \ }stand for the all order corrections. An
important observation in evaluating the unknown functions $g^{(k)}({\bf {x}}%
_{b},{\bf {x}}_{a};s)$ is that the curly bracket above satisfies the
Schr\"{o}dinger-like equation 
\begin{equation}
\left[ \hat{{\cal H}}_{E}\left( -i\hbar \partial _{{\bf {x}}_{b}},{\bf {x}}%
_{b};s\right) -i\hbar \partial _{s}\right] K_{{\rm all}}({\bf {x}}%
_{b},s_{b}\mid {\bf {x}}_{a},s_{a})=-i\hbar \delta \left( s_{b}-s_{a}\right)
\delta ^{3}\left( {\bf {x}}-{\bf {x}}_{a}\right) ,  \label{3}
\end{equation}
where $\hat{{\cal H}}_{E}$ is the Hamilton operator 
\begin{equation}
\hat{{\cal H}}_{E}\left( \hat{{\bf p}},{\bf x};s\right) =\left( \hat{{\bf p}}%
-e/c{\bf A}({\bf x,}s)\right) ^{2}/2M+[2EV({\bf x,}s)-V^{2}({\bf x,}%
s)]/2Mc^{2}
\end{equation}
with ${\bf \hat{p}}=-i\hbar \partial _{{\bf {x}}}$. Since the boundary
condition given in Eq. (\ref{sc}) and the limiting property is $K_{{\rm all}%
}({\bf {x}}_{b},s_{a}\mid {\bf {x}}_{a},s_{a})=\delta ^{3}\left( {\bf {x}}%
_{b}-{\bf {x}}_{a}\right) $ for the pseudopropagator $K_{{\rm all}}$, we
obtain the relations 
\begin{equation}
\left\{ 
\begin{array}{l}
g^{(0)}({\bf {x}}_{b},s_{b}\mid {\bf {x}}_{a},s_{a})=1 \\ 
g^{(k)}({\bf {x}}_{b},s_{a}\mid {\bf {x}}_{a},s_{a})=0,\quad k\geq 1
\end{array}
\right. .  \label{d}
\end{equation}
To go further, let us insert the pseudopropagator $K_{{\rm all}}$ in Eq. (%
\ref{2}) into Eq. (\ref{3}). Three equalities arise from this operation.
They are

\begin{equation}
\frac{1}{2M}\left[ \partial _{x^{i}}{\cal A}_{E}({\bf {x}},s\mid {\bf {x}}%
_{a},s_{a})-e/cA^{i}{\bf (x,}s{\bf )}\right] ^{2}+V_{E}({\bf x,}s)=\partial
_{s}{\cal A}_{E}({\bf {x}},s\mid {\bf {x}}_{a},s_{a}),  \label{a}
\end{equation}
where $V_{E}({\bf x,}s)$ is defined as $[2EV({\bf x,}s)-V^{2}({\bf x,}%
s)]/2Mc^{2}$,
\[
\partial _{s}D_{E}^{1/2}({\bf {x}},s\mid {\bf {x}}_{a},s_{a})
\]
\begin{equation}
=-\frac{1}{2M}\left[ 2(\partial _{x^{i}}D_{E}^{1/2})\left( \partial _{x^{i}}%
{\cal A}_{E}-e/cA^{i}\right) +D_{E}^{1/2}\partial _{x^{i}}\left( \partial
_{x^{i}}{\cal A}_{E}-e/cA^{i}\right) \right] ,  \label{b}
\end{equation}
and the iterative equation 
\[
\left\{ \partial _{s}+\frac{1}{M}\left[ \partial _{x^{i}}{\cal A}_{E}({\bf {x%
}},s\mid {\bf {x}}_{a},s_{a})-e/cA^{i}{\bf (x,}s{\bf )}\right] \partial
_{x^{i}}\right\} g^{(k)}({\bf {x}},s\mid {\bf {x}}_{a},s_{a})
\]
\begin{equation}
=\hat{O}_{E}({\bf {x}},s\mid {\bf {x}}_{a},s_{a})g^{(k-1)}({\bf {x}},s\mid 
{\bf {x}}_{a},s_{a}).  \label{c}
\end{equation}
Here the action of the operator $\hat{O}_{E}$ is defined as 
\begin{equation}
\hat{O}_{E}({\bf {x}},s\mid {\bf {x}}_{a},s_{a})g^{(k-1)}({\bf {x}},s\mid 
{\bf {x}}_{a},s_{a})=D_{E}^{-1/2}({\bf {x}},s\mid {\bf {x}}%
_{a},s_{a})\partial _{i}^{2}\left[ D_{E}^{1/2}g^{(k-1)}\right] .
\end{equation}
Eqs. (\ref{a}) and (\ref{b}) is just the Hamilton-Jacobi equation and
continuity equation, respectively. Eq. (\ref{c}) will provide us the
information of each order. To solve it, we note that the middle bracket in
Eq. (\ref{c}) precisely satisfies the first of the equation: 
\begin{equation}
\frac{d}{ds}x^{i}(s)=\left. \frac{1}{M}\left[ \partial _{x^{i}}{\cal A}_{E}(%
{\bf {x}},s\mid {\bf {x}}_{a},s_{a})-\frac{e}{c}A^{i}{\bf (x,}s{\bf )}\right]
\right| _{{\bf x}={\bf x}(s)}
\end{equation}
if we identify $g^{(k)}({\bf {x}},s\mid {\bf {x}}_{a},s_{a})=g^{(k)}({\bf {x(%
}}s;{\bf {x}}_{b},s_{b}\mid {\bf {x}}_{a},s_{a}),s\mid {\bf {x}}_{a},s_{a}).$
The left-hand side of Eq. (\ref{c}) now equals the total derivative 
\begin{equation}
\frac{d}{ds}g^{(k)}({\bf {x(}}s;{\bf {x}}_{b},s_{b}\mid {\bf {x}}%
_{a},s_{a}),s\mid {\bf {x}}_{a},s_{a})={\rm left\;hand\;side\;of\;Eq}.\;(\ref
{c}).
\end{equation}
From this, it is easy to find the explicitly solution, subject the condition
in Eq. (\ref{d}), 
\[
g^{(k)}({\bf {x}}_{b},s_{b}\mid {\bf {x}}_{a},s_{a})
\]
\begin{equation}
=\frac{i}{2M}\left. \int_{s_{a}}^{s_{b}}ds\hat{O}_{E}({\bf {x}},s\mid {\bf {x%
}}_{a},s_{a})g^{(k-1)}({\bf {x}},s\mid {\bf {x}}_{a},s_{a})\right| _{{\bf {x(%
}}s;{\bf {x}}_{b},s_{b}\mid {\bf {x}}_{a},s_{a})}.
\end{equation}

It is useful to introduce a {\it pseudotime-ordering operator }which orders
the pseudotimes successively 
\begin{equation}
\hat{T}_{s}\left( \hat{O}_{E}(s_{n})\hat{O}_{E}(s_{n-1})\cdots \hat{O}%
_{E}(s_{1})\right) \equiv \hat{O}_{E}(s_{i_{n}})\hat{O}_{E}(s_{i_{n-1}})%
\cdots \hat{O}_{E}(s_{i_{1}}),
\end{equation}
where $s_{i_{n}},\cdots ,s_{i_{1}}$ are the pseudotimes $s_{n},\cdots ,s_{1}$
relabeled in the causal order, so that 
\begin{equation}
s_{i_{n}}\geq s_{i_{n-1}}\geq \cdots \geq s_{i_{1}}.
\end{equation}
With this formal operator, the expansion can be rewritten in a more compact
form and is given by 
\[
G_{{\rm all}}({\bf {x}}_{b},{\bf {x}}_{a};E)=\frac{\hbar }{2Mc}\frac{1}{%
(2\pi \hbar i)^{D/2}}\int_{s_{a}}^{\infty }ds_{b}e^{\left[ \frac{i}{\hbar }{%
\varepsilon }\left( s_{b}-s_{a}\right) \right] }D_{E}^{1/2}({\bf {x}}%
_{b},s_{b}\mid {\bf {x}}_{a},s_{a}) 
\]
\begin{equation}
\times e^{\left[ \frac{i}{\hbar }{\cal A}_{E}({\bf {x}}_{b},s_{b}\mid {\bf {x%
}}_{a},s_{a})\right] }\hat{T}_{s}\exp \left\{ \frac{i\hbar }{2M}%
\int_{s_{a}}^{s_{b}}ds\hat{O}_{E}({\bf {x}},s\mid {\bf {x}}%
_{a},s_{a})\right\} .
\end{equation}
Considering the phase change coming from the conjugate points of classical
orbits \cite{2.8,3} and the various classical trajectories joining the $(%
{\bf {x}}_{a},s_{a})$ and $({\bf {x}}_{b},s_{b})$ in the generic space-time,
we have 
\[
G_{{\rm all}}({\bf {x}}_{b},{\bf {x}}_{a};E)=\frac{\hbar /2Mc}{(2\pi \hbar
i)^{D/2}}\sum_{{\rm classical\;orbit}}\int_{s_{a}}^{\infty }ds_{b}e^{\left[ 
\frac{i}{\hbar }{\varepsilon }\left( s_{b}-s_{a}\right) \right] }D_{E}^{1/2}(%
{\bf {x}}_{b},s_{b}\mid {\bf {x}}_{a},s_{a}) 
\]
\begin{equation}
\times e^{\left[ \frac{i}{\hbar }{\cal A}_{E}({\bf {x}}_{b},s_{b}\mid {\bf {x%
}}_{a},s_{a})-i\mu \pi /2\right] }\hat{T}_{s}\exp \left\{ \frac{i\hbar }{2M}%
\int_{s_{a}}^{s_{b}}ds\hat{O}_{E}({\bf {x}},s\mid {\bf {x}}%
_{a},s_{a})\right\} .  \label{all}
\end{equation}
This result relates to the operation of Van Vleck determinant. Since the
summation of the perturbation series should converges to the fixed-energy
amplitude, we have a new point of view for quantization which just relates
to the topology and the summation of classical orbits. Contrary to the
Feynman's path integral, where many nonclassical paths are summed, we just
need to consider the physical paths here. We believe that classical orbits
are the most fundamental factor for quantization. This idea was exploited by
Bohr's in 1913 by postulating the famous rule that only a countable number
of orbits are allowed by the quantum condition $\oint pdq=nh\;(n\in N)$.
However, this ``old'' quantum mechanics is not sufficient, because it does
not consider the topology of the classical orbits \cite{4}.

The idea of summing the classical orbits for quantization can be found in
the trace formula \cite{3}. The energy spectra produced by taking the trace
of the non-relativistic version of Eq. (\ref{all}) even in the lowest order
approximation are highly accurate. Another evidence for quantization via
classical orbits comes from gauge transformation. The well-known space-time
technique in path integral is a gauge transformation \cite{2.5}. It stands
for the over- summation of the classical orbits, think for instance of gauge
fixing techniques in gauge field theory where divergence arises for over
counting the gauge fields when we perform the sum over gauge fields using
path integral.

Eq. (\ref{all}) is also suitable for the non-relativistic systems. This is
achieved by replacing $V_{E}({\bf x,}s)$ and $\varepsilon $ with $V({\bf x,}%
s)$ and $E$, respectively. Numerous physical problems such as tunneling
effect, quantization of classical chaotic systems, and the systems related
to the semiclassical quantization require higher order corrections. We hope
that our result may offer a useful tool for performing these calculations.
Particularly, the formula given in Eq. (\ref{all}) may provide systems
failing to be quantized using Feynman's method an alternate way of
quantization via classical orbits.

Details about the calculations discussed in this letter and illustrative
examples will be presented elsewhere. We expect that the formula in Eq. (\ref
{all}) provide detailed microscopic information via the classical orbits.

\centerline{ACKNOWLEDGMENTS} \centerline{The author is greatful to Prof. M.
C. Chang for helpful discusions.} 


\newpage

\begin{references}
\bibitem{0}  J. H. Van Vleck, Proc. Natn. Acad. Sci. {\bf 14}, 178 (1928).

\bibitem{0.5}  M. V. Berry and K. E. Mount, Rep. Prog. Phys. {\bf 35}, 315
(1972).

\bibitem{1}  H. Kleinert, Phys. Lett. {\bf A 212}, 15 (1996).

\bibitem{2}  H. Kleinert, {\it {Path Integrals in Quantum Mechanics,
Statistics and Polymer Physics}}, World Scientific, Singapore, 1995.

\bibitem{2.5}  D. H. Lin, J. Phys. {\bf A 30} 3201 (1997); {\bf A 30} 4365
(1997); {\bf A 31} 4785 (1998); {\bf A 31} 7577 (1998); hep-th/9708144 to
appear in J. Math. Phys.; hep-th/9709152.

\bibitem{2.6}  K. Fujikawa, Prog. Theor. Phys. {\bf 96, }863 (1996)
(hep-th/9609029); (hep-th/9608052).

\bibitem{lin}  D. H. Lin, {\it Path Integral on Relativistic Spinless
Potential Problems}, talk given at the sixth International Conference on
Path-Integrals from peV to TeV 50 Years from Feynman's Paper, Florence,
Italy, 25-29 August 1998, to appear in the Proceedings.

\bibitem{2.7}  H. Kleinert and D. H. Lin, {\it Relativistic Trace Formula
for Bound States in Terms of Classical Periodic Orbits}, quant-ph/9807068.

\bibitem{mr}  M. Roncadelli, Phys. Rev. Lett. {\bf 72,} 1145 (1994).

\bibitem{2.8}  J. B. Keller, Ann. Phys. {\bf 4}, 180 (1958).

\bibitem{3}  M. Gutzwiller, {\it Chaos in Classical and Quantum Mechanics},
Springer, Berlin, 1990.

\bibitem{4}  D. Wintgen, K. Richter, and G. Tanner, CHAOS {\bf 2}, 19 (1992).
\end{references}
\end{document}